\documentclass[a4paper,fleqn,usenatbib]{mnras}

\usepackage[T1]{fontenc}
\usepackage{ae,aecompl}

\usepackage{anyfontsize}
\usepackage{amsmath}
\usepackage{amsfonts}
\usepackage{amsbsy}

\usepackage{newtxtext}
\usepackage{newtxmath}

\usepackage{tensor}
\usepackage{mathrsfs}
\usepackage{bm}

\usepackage[utf8]{inputenc}

\usepackage{graphicx}
\usepackage{epsfig}
\usepackage{epstopdf}

\usepackage[usenames,dvipsnames]{xcolor}
\usepackage{hyperref}

\newcommand{\be}{\begin{equation}}
\newcommand{\ee}{\end{equation}}

\usepackage{graphicx}	
\usepackage{amsmath}	

\title[Drifting through the medium]{Drifting through the medium: \\kicks and self-propulsion of binaries within accretion disks and other environments}

\author[V. Cardoso and C. F. B. Macedo]{
Vitor Cardoso,$^{1,2}$\thanks{E-mail: vitor.cardoso@ist.utl.pt}
and Caio F. B. Macedo,$^{3}$\thanks{E-mail: caiomacedo@ufpa.br}
\\
$^{1}$CENTRA, Departamento de Física, Instituto Superior Técnico – IST,Universidade de Lisboa – UL, Avenida Rovisco Pais 1, 1049 Lisboa, Portugal\\
$^{2}$Waseda Institute for Advanced Study (WIAS), Waseda University, Shinjuku, Tokyo 169-8050, Japan\\
$^{3}$Faculdade de Física, Universidade Federal do Par\'a, Salin\'opolis, Par\'a, 68721-000 Brazil
}

\begin{document}
\label{firstpage}
\pagerange{\pageref{firstpage}--\pageref{lastpage}}
\maketitle

\begin{abstract}
Compact binaries are within the reach of gravitational and electromagnetic wave detectors, and are important for our understanding of astrophysical environments and the composition of compact objects. There is a vast body of work devoted to the evolution of such binaries in background media, such as in common-envelope evolution, accretion disks and dark matter mini-spikes. Here, we explore further gravitationally-bound binaries evolving within an environment. We show that dissipative effects such as gravitational drag and accretion impart a momentum to the center of mass of asymmetric binaries.
We numerically evolve the binaries in a Newtonian setup and show that, depending on the medium density, the center of mass can accelerate to high 
speeds -- in some cases $300\, {\rm km/s}$ or more -- during inspiral, with potentially observable signatures.  
Our numerical results are fully consistent with an \textit{analytical} result for the CM evolution at first order in the medium density. 
\end{abstract}




\section{Introduction}
\label{sec:int}

Self-propulsion with the use of the surrounding environment is ubiquitous in biology, and is used by micro-organisms such as {\it Escherichia Coli}~\citep{amjphys:1977,PhysRevLett.58.2051}.
In astrophysics, the environment is thought as a dissipative medium which slows planets, stars and black holes down, through accretion and gravitational drag \citep{Chandrasekhar:1943ys,Ostriker:1998fa,Shapiro:1983du}. However, binary systems behave in similar ways to living micro-organisms: periodic motion of an asymmetric system leads eventually to motion of its center-of-mass (CM). 

Thus far, the CM motion of compact binaries has been well-studied in the context of gravitational-wave emission. The linear momentum carried by these waves imparts a net ``kick'' to the CM~\citep{Gonzalez:2006md,Campanelli:2007cga,Centrella:2010mx}. This kick is mostly acquired during the last stages of inspiral and has important consequences for the astrophysics of such objects or their host galaxies. Kicks of $\gtrsim 1000\,{\rm km/s}$ can exceed the escape velocity of most galaxies, thus ejecting black holes from their hosts~\citep{Merritt:2004xa,Volonteri_2010,Gerosa:2014gja,Sesana:2007zk}; smaller kicks are still able to displace black holes from the galaxy core during long timescales~\citep{Gualandris:2007nm,Komossa:2008as}. Such motion leads to different electromagnetic and gravitational-wave signatures, and is an active field of research~\citep{Komossa:2012cy,Gerosa:2016vip}.

Here, we are instead interested in the CM motion induced by interaction with the environment, in particular through dynamical friction. Such interaction has been explored, and is known to drive the inspiral faster, providing clear smoking-gun signatures of nontrivial astrophysical environments via gravitational-wave tracking of the rate of inspiral~\citep{Macedo:2013qea,Barausse:2014tra,Vicente:2019ilr,Cardoso:2019rou}. What happens to the CM of the binary as it evolves under dynamical friction and accretion? This elementary problem -- which can be addressed with purely Newtonian physics -- is the focus of this work. It is specially interesting in the context of supermassive binaries evolving in accretion disks which have high densities (as large as $10^{-6} - 100\,{\rm kg}/{\rm m}^3$ for thick and thin accretion  disks,  respectively~\citep{Barausse:2014tra}) or for coalescing black holes formed via dynamical fragmentation of a very massive star undergoing gravitational collapse, leading to a binary evolving in a medium with density as high as $10^{10}\,{\rm kg}/{\rm m}^3$  or higher~\citep{Loeb:2016fzn,Reisswig:2013sqa}. 
This is also important in the common envelope stage of binary stars, in which the stars orbit each other inside a common medium envelope, where densities are usually smaller then $10^{-2}\,{\rm kg/m^3}$~\citep{2013A&ARv..21...59I,Passy_2011}. In fact, one can question whether this phenomena is of relevance even in the context of dark matter physics, for sufficiently large-density environments, as in dark matter mini-spike scenarios~\citep{Eda:2013gg,Kavanagh:2020cfn}.

\section{Binary configurations moving in a medium}
We consider a binary composed by two compact objects, of masses $m_i\,(i=1,2)$ a distance $r$ apart, evolving under their mutual gravitational attraction, and under additional non-conservative forces $\mathbf{F}_i$.
The equations describing the motion of the particles are
\begin{equation}
m_{i}\ddot{\mathbf{r}}_{i}+\dot{m}_{i}\dot{\mathbf{r}}_{i}=\pm \frac{G m_1 m_2}{r^3}\mathbf{r}+\mathbf{F}_i,
\label{eq:bin}
\end{equation}
where $\mathbf{r}_i$ is the coordinate position of each particle, $\mathbf{r}=\mathbf{r}_2-\mathbf{r}_1$. Since we are interested in the net effect that the media may have in the CM motion, we focus on binaries with vanishing relative CM velocity~\citep{Pani:2015qhr}.

In a non-trivial environment, there are different forces acting on the objects. In this work, we will be concerned with two dissipative forces:

\noindent \textbf{\textit{Accretion}}.  
Accretion of surrounding matter effectively changes the motion by increasing the mass of the individual particles. Accretion can be modelled as an effective force by~(see e.g. \cite{Macedo:2013qea})
\begin{equation}
\mathbf{F}_{{\rm a},i}=-\dot{m}_i\dot{\mathbf{r}}_i\,,
\end{equation}
where $\dot{m}_i=\zeta_i\rho v_i$, where $\rho$ is the density of the medium, $\zeta_i$ is the accretion cross section related to the $i$-object and $v_i$ their speed relative to the medium. We can further assume that the medium is such that $v_i=|\dot{\mathbf{r}}|$. The accretion cross section is highly dependent on the model and on the compact object. We can, however, estimate the importance of accretion by considering a simple model with $\zeta_i=\pi {R_{{\rm eff},i}}^2$, where ${R_{{\rm eff},i}}$ is the effective capture radius of the accreting compact object~\citep{Shapiro:1983du}. For compact objects such as black holes, for instance, considering nonrelativistic accretion, we have $R_{{\rm eff},i}\sim R_{i} c/v_i$, with $R_i$ being the effective size of the compact object. Therefore, we shall consider the effective force due to accretion to be
\be
\mathbf{F}_{{\rm a},i}=-\frac{\pi c^2\rho R_i^2}{v_i}\dot{\mathbf{r}_i}.
\ee
It is worth to note that for black holes $R_i\sim G m_i/c^2$, and so we have $|F_i|=\pi G^2m_i^2\rho/v_i$. We shall use this expression to estimate the importance of accretion in the binary evolution.

\noindent \textbf{\textit{Gravitational drag}}. A mass moving in a medium generates a gravitationally-induced density wake that provides a gravitational drag force. This is known as dynamical friction and has been studied in many different scenarios, since the seminal work by~\cite{Chandrasekhar:1943ys}. When the physical dimension of the medium is very large in comparison with the typical size of the moving bodies, one can treat the problem as a constant density medium. Assuming that the binary separation is large enough to neglect interactions with the wakes of each moving body, the gravitational drag force can be described by
\be
\mathbf{F}_{{\rm d},i}=-G^2 m_i^{2} \rho I (v_i) \dot{\mathbf{r}}_i\,,
\label{key}
\ee
where $I (v_i)$ depends on the model for the fluid. 

For fluid-like media with pressure, considering a linear motion,~\cite{Ostriker:1998fa} found that,
\be
I(v_i)=\frac{2 \pi}{v_i^{3}}\left\{\begin{array}{ll}{\ln \left(\frac{1+{\cal M}_i}{1-{\cal M}_i} e^{-2 {\cal M}_i}\right),} & {{\cal M}_i<1} \\ {\ln \left(\Lambda-\frac{\Lambda}{{\cal M}_i^{2}}\right),} & {{\cal M}_i>1}\end{array}\right.\,,
\label{eq:dyn}
\ee
where ${\cal M}_i=v_i/c_s$ is the Mach number, with $c_s$ being the sound speed in the medium. The value of $\Lambda$ depends on the time the interaction takes place, but it can be adjusted to fit special particular cases of motion (e.g., \cite{Kim:2007zb} chosen $\Lambda$ to fit circular orbits). As can be seen in the expression~\eqref{eq:dyn}, there is an enhancement in the dynamical friction for velocities $v_i\sim c_s$. As such, since we shall deal with binary configurations, it is useful to remind ourselves that binary orbital velocities are typically
\begin{equation}
v\sim 1.21\times 10^{-4} (1+m_1/m_2)^{1/2}\left(\frac{m_1}{10 M_\odot}\right)^{1/2}\left(\frac{10^7 {\rm km}}{a}\right)^{1/2}c\,,
\end{equation}
which can easily be supersonic in many astrophysically relevant scenarios (e.g.,the speed of sound in our sun is $550.000{\rm m/s}\sim 1.82\times10^{-6}c$). This also happens for the binary evolution in common envelopes~\citep{Iben:1993yi}. Supersonic motion usually experiences a stronger drag force, mainly in the transitions between sub and supersonic motions~\citep{Ostriker:1998fa,Macedo:2013qea}.

On the other hand, for pressureless media,~\cite{Chandrasekhar:1943ys} found that
\be
I(v_i)=\frac{4\pi\lambda}{v_i^3}\left[{\rm erf}(v_i/(\sqrt{2}\sigma))-\frac{2 v_i}{\sqrt{2\pi}\sigma}e^{v_i^2/(2\sigma^2)}\right],
\label{eq:chandra}
\ee
with $\sigma$ being the dispersion of the matter Maxwellian velocity distribution, and $\lambda$ is the Coulomb logarithm~\citep{RevModPhys.21.383,Binney}. In this paper, we shall take $\lambda=20$, for simplicity.

\begin{figure}
	\includegraphics[width=\linewidth]{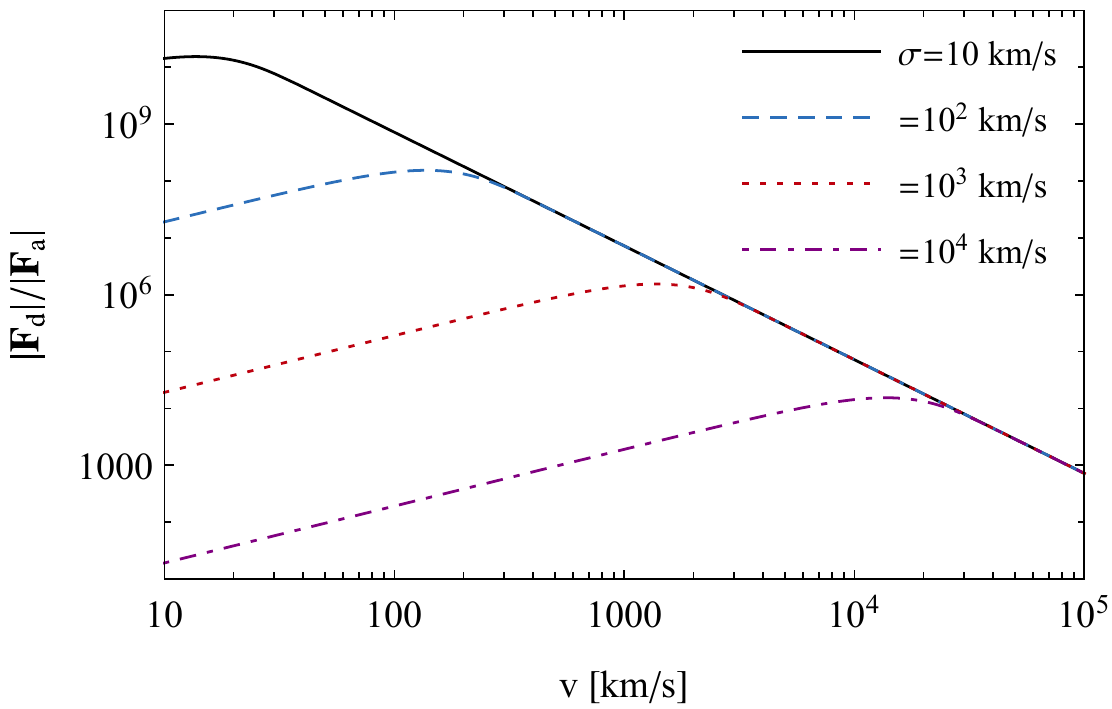}
	\caption{Ratio between the moduli of the dissipative force induced by dynamical friction, according to the Chandrasekhar model, and collisionless accretion. We can see that dynamical friction dominates. A similar conclusion holds for the supersonic Bondi-Hoyle accretion model~\citep{Macedo:2013qea}.}
	\label{fig:comp}
\end{figure}
Due to the simplicity of the above expressions for the dissipative forces related to accretion and dynamical friction, we can easily compare their relative importance in binary motion. In Fig.~\ref{fig:comp} we plot the ratio of the two forces as function of the velocity for the Chandrasekhar model, for different values of $\sigma$. In general, accretion is subdominant for the most part of the parameter space. A similar conclusion holds for the Ostriker model. As such, we shall neglect accretion and deal only with dynamical friction for our computations.

For simplicity, we focus primarily on the Chandrasekhar model to quantify dynamical friction [given by Eq.~\eqref{eq:chandra}]. However, some of the most important results generalize to
collisional dynamical friction. We quantify this in Appendix~\ref{app:collisional}, where we show that the Ostriker model provides also similar, large CM velocities. Therefore, we expect that the conclusions presented here can be extended to other scenarios. In addition, we assume that the timescale of the orbital evolution is longer than the replenishment timescale of the wake behind the orbiting bodies. This assumption is necessary because otherwise the individual wakes would interfere with each other modifying the dynamical friction expressions~\citep{Kim:2008ab,Baruteau:2010sm}. Note, however, that the additional force component due to the companion wake decreases considerably for ${\cal M}\gg1$, as shown by \cite{Kim:2008ab}.

Since we are interested in tracking the evolution of the CM under the non-conservative forces, it is useful to introduce the equations describing $(\mathbf{r},\mathbf{R})$ instead of $(\mathbf{r}_1,\mathbf{r}_2)$, with
\be
\mathbf{R}=\frac{m_1\mathbf{r}_1+m_2\mathbf{r}_2}{m_1+m_2},
\ee
begin the CM position of the binary. We shall denote the capital letters $(\mathbf{R},\mathbf{V})$ for the position and velocity of the center of mass and $(\mathbf{r},\mathbf{v})$ for the binary separation quantities.
Writing the non-conservative force in the form
\begin{equation}
\mathbf{F}_i=-G^2m_i^2\rho I_i\dot{\mathbf{r}}_i\,,
\end{equation}
where $I_i\equiv I(v_i)$ represents functions of the $i$-particle speed\footnote{Note that these can also be written in function of the center of mass and binary separation, as can be seen in App.~\ref{app:equations}.}, given by Eq.~\eqref{eq:chandra}, we obtain
\begin{align}
\ddot{\mathbf{R}}&=-\frac{G^2 M \rho }{(q+1)^2}(I_1+q^2I_2)\dot{\mathbf{R}} +\frac{G^2 M \rho q }{(q+1)^3}(I_1-q I_2)\dot{\mathbf{r}},\label{eq:center}\\
\ddot{\mathbf{r}}&=-\frac{G^2 M \rho q  }{(q+1)^2}(I_1+I_2)\dot{\mathbf{r}}+\frac{G^2 M \rho }{q+1}(I_1-q I_2)\dot{\mathbf{R}}-\frac{GM}{r^3}\mathbf{r}\,.\label{eq:disp}
\end{align}
Here, $M=m_1+m_2$ is the total mass, while 
\be
q=\frac{m_2}{m_1}\,,
\ee
is the binary the mass ratio (which we shall take $q\geq 1$ without loss of generality).

When the binary is symmetric, $m_1=m_2$, we have that $(I_1-q I_2)$ vanishes identically, as $I_1=I_2$, and the equations decouple into
\begin{align}
\ddot{\mathbf{R}}&=-\frac{1}{2} G^2 M \rho I_1\dot{\mathbf{R}},\label{eq:cmd}\\
\ddot{\mathbf{r}}&=-\frac{1}{2} G^2 M \rho I_1\dot{\mathbf{r}}-\frac{GM}{r^3}\mathbf{r}\,.\label{eq:bid}
\end{align}
Equation~\eqref{eq:cmd} essentially tells us that the CM tends to remain at rest relative to the medium. The evolution of the binary, therefore, has no impact in boosting the CM, providing only a dragging force. The binary suffers from an effective drag force, and by defining the angular momentum per unit of mass as as $L=r^2\dot{\varphi}(t)$, with $\varphi$ being the standard angular coordinate in a polar decomposition~(see, e.g., ~\citep{Macedo:2013qea}), we obtain that
\be
\dot{L}=-\frac{1}{2} G^2 M \rho I_1 L.
\label{eq:angularm}
\ee
Therefore, the angular momentum of the binary decreases in time. The timescale for the coalescence depends on the form of $I_1$, which evolves with the binary. We should note that inside the definition of $I_1$ there are factors that depend on $V$ and $v$ (and, therefore, on the angular momentum), so we cannot straightforwardly integrate Eq.~\eqref{eq:angularm}.

\begin{figure}
	\begin{tabular}{c}
		\includegraphics[width=\linewidth]{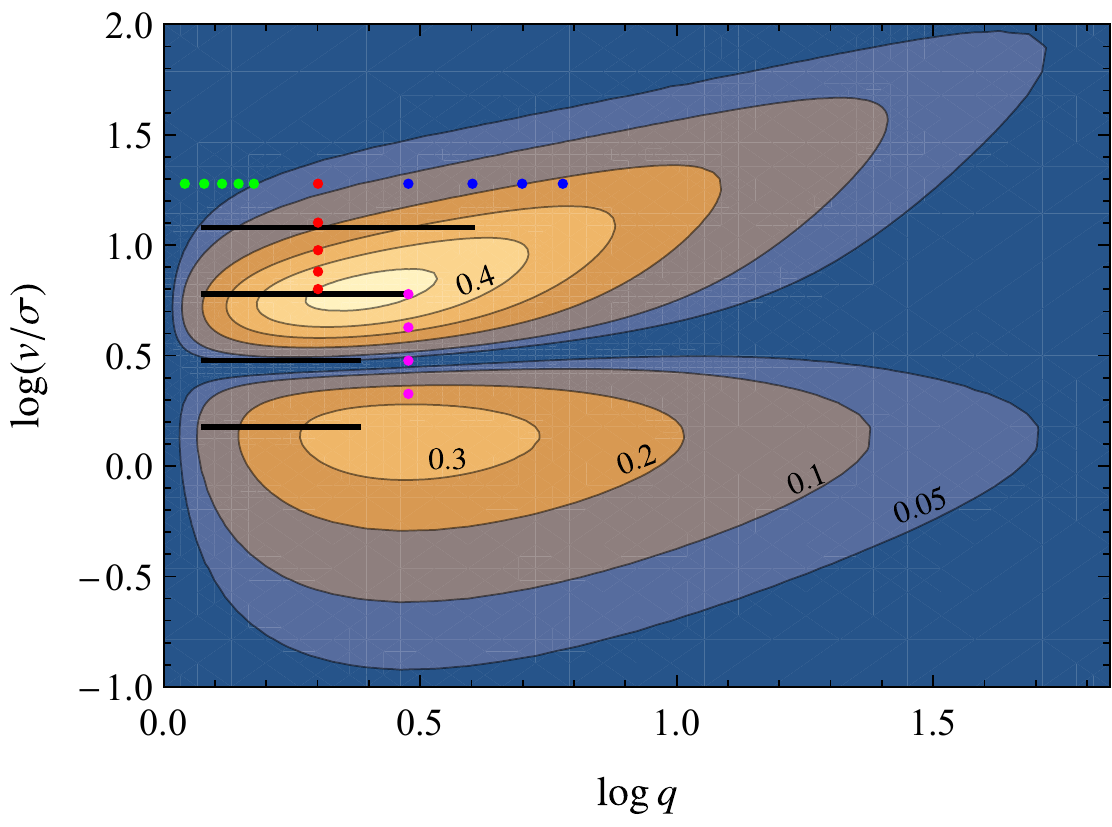}\\
		\includegraphics[width=\linewidth]{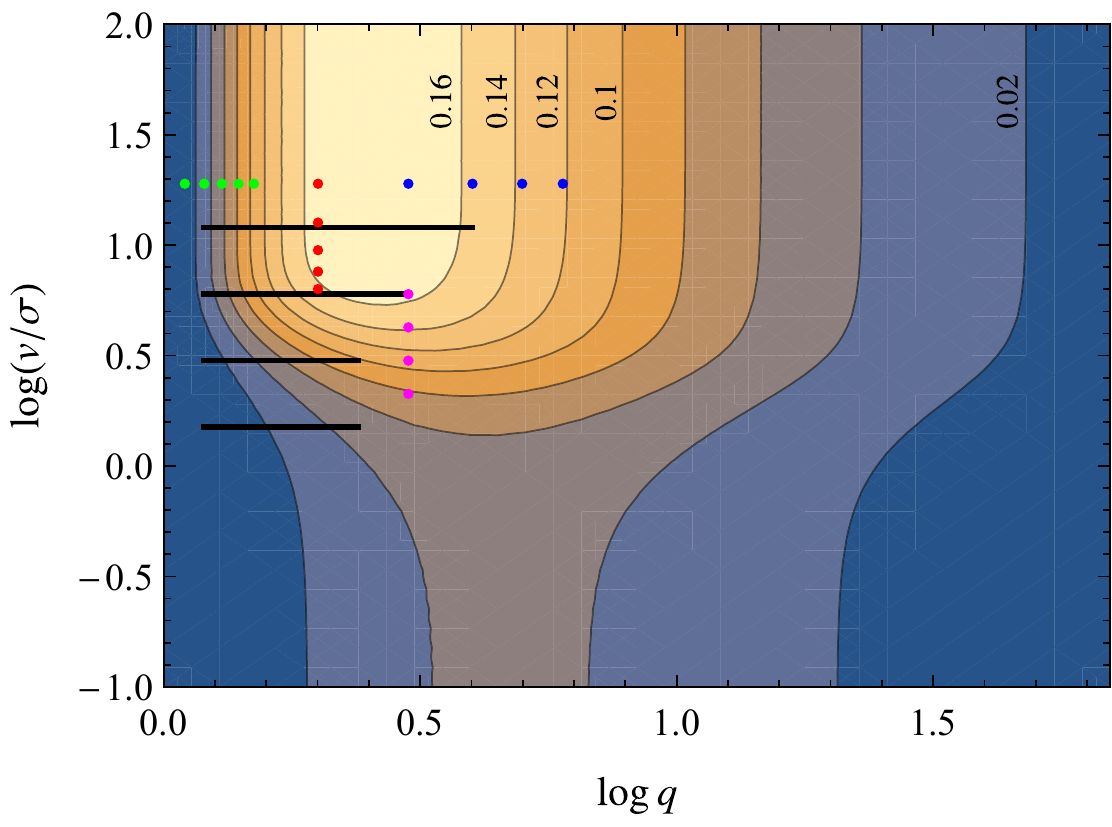}
	\end{tabular}
	\caption{Predictors for the CM boost. {\bf Top:} Absolute value of the force responsible for the CM boost as function of the binary speed $v$ (normalized by the dispersion) and the mass ratio $q=m_2/m_1$. The dots mark the location of the configurations used to numerically compute the evolution of the binaries (see Sec.~\ref{sec:num}).
		{\bf Bottom:} $\eta_1 \eta_2$ for different mass ratio and binary speed. The lighter-colored region indicates binaries that can potentially have considerable kicks. The dots and horizontal lines mark the configurations computed numerically in Sec.~\ref{sec:num}.}
	\label{fig:force}
\end{figure}

For asymmetric binaries, however, there is an additional force proportional to $\sim (I_1-q I_2)$, which will evolve with the binary. This term is responsible for boosting the CM. We can analyze the parameter space in which the coupling term of Eq.~\eqref{eq:center} is important. Consider the Chandrasekhar model, $|\dot{\mathbf{r}}|=v$, and (for the moment, we assume $|\dot{\mathbf{R}}|=0$)
\begin{align}
v_1=\frac{q}{q+1}v,~v_2=\frac{1}{q+1}v\,.
\end{align}
We can normalize the force of the center of mass by the factor
\begin{equation}
\alpha\equiv \frac{G^2 M^2 \rho }{(q+1) \sigma ^2}\,,\label{eq:scale}
\end{equation}
giving a dimensionless estimate of the force acting in the CM 
(this value also coincides with the dimensionless ``acceleration'', i.e., the force divided by $M$). The top panel of Fig.~\ref{fig:force} shows the force as function of the binary parameters. We can see that there is a region in the ($q$,$v/\sigma$), lighter regions in the figure) plane that maximizes the force. Binary configurations living in that region will potentially experience a large CM boost due to the dissipative forces. The dots and lines indicate the configurations explored in Sec.~\ref{sec:num}.

\begin{figure*}
	\includegraphics[width=\linewidth]{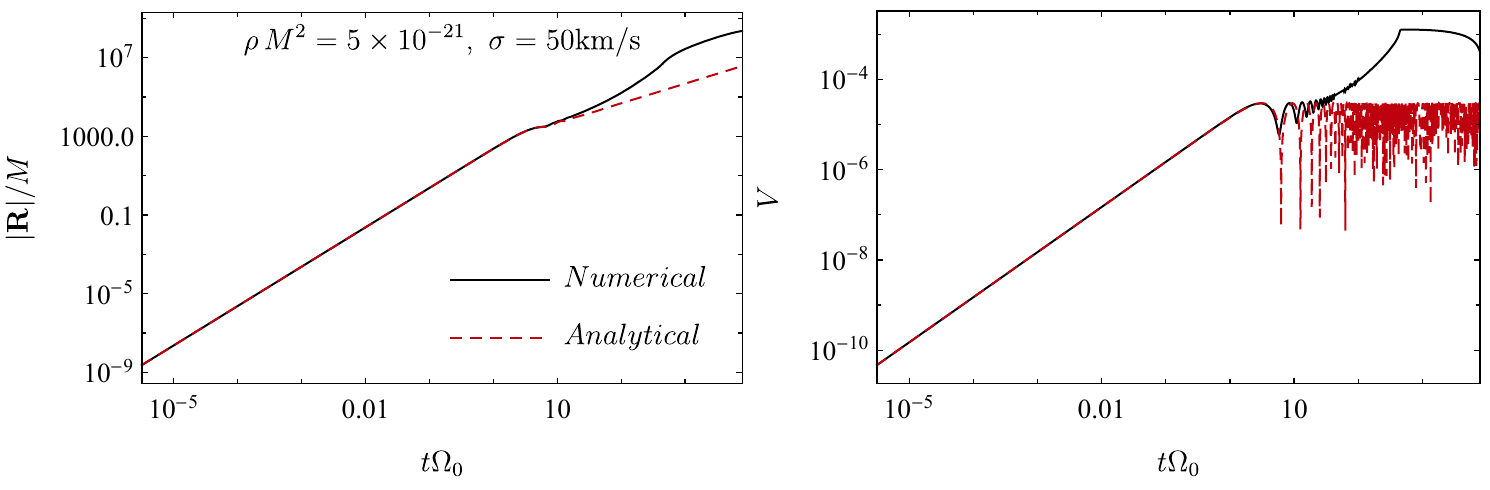}
	\includegraphics[width=\linewidth]{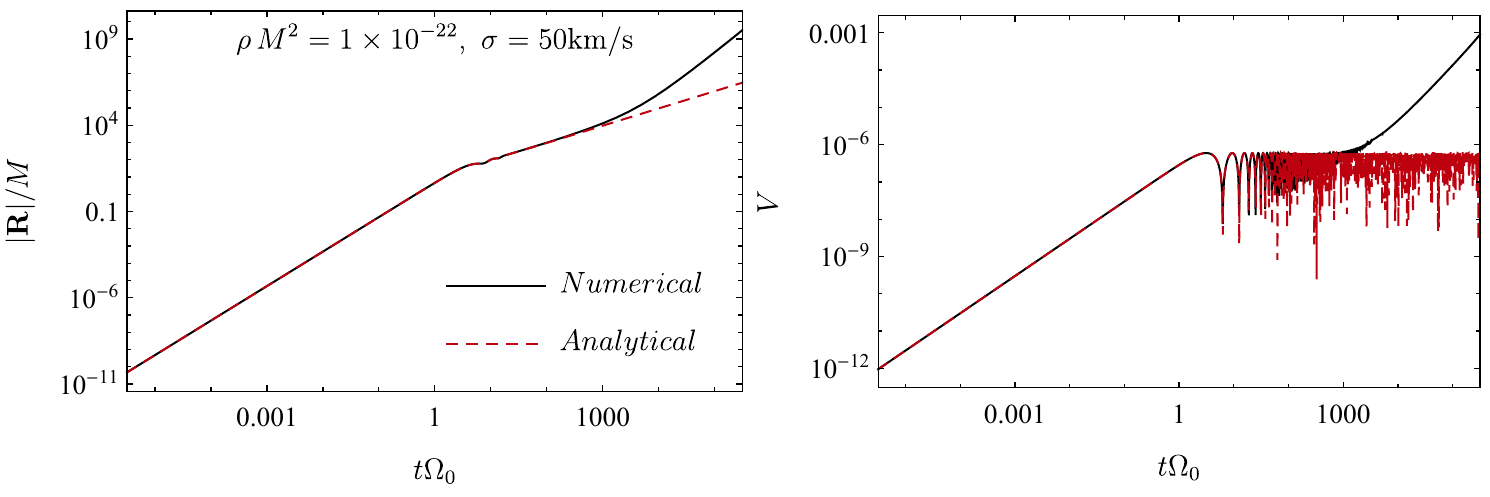}
	\caption{Comparison between the numerical integrated solution of Eqs.~\eqref{eq:center}--\eqref{eq:disp} and the analytical solution given by Eq.~\eqref{eq:low_density}. The horizontal axis is time scaled in units of the inverse initial orbital frequency; it is thus a measure of the number of orbits by the binary. The orbital radius is initially $a=10^5 M$ and the mass ratio is $q=2$. For smaller densities, the agreement is better during longer timescales, as can be seen comparing the top and lower panels. We have used $G=c=1$ [note the conversion factor given by Eq.~\eqref{eq:conversion_m}].}
	\label{fig:analytical}
\end{figure*}
The top panel of Fig.~\ref{fig:force} is a useful guide to search for binary configurations with potential environmental kicks. Note, however, that it should be taken with care, as it was built under the assumption that the CM is at rest and that the binary is in an initially circular orbit. Moreover, it only tells us about the magnitude of the force, but the analysis of the kick is much more intricate. For instance, higher-density profiles in the lighter-colored region (i.e., corresponding to large forces acting on the CM, as shown in Fig.~\ref{fig:force}) could generate orbits that are very short-lived, in the sense that they last for a small number of cycles before merging; in such a case, the CM would not have had time to accumulate a considerable speed. Additionally, one can start in the lighter-colored region but, since the binary speed $v$ increases (in fact, the orbit could become highly eccentric), the initial point could actually move in the plane and quickly migrate to other regions in the plane. 

From the above reasoning, one could expect that there exist other important ``predictors'' of the CM motion. From the qualitative picture we have three forces that are potentially relevant: i) The one responsible for the CM boost, ii) the CM drag force, and iii) the binary drag force. We can analyze these in terms of the effective acceleration each provide (force/effective mass). As discussed above, if the gravitational drag in the binary is relevant, the CM may not have considerable time to acquire speed. As such, it is natural to consider the ratio between the CM boost and the binary dragging. We have
\be
\eta_1=\frac{1}{q+1}\frac{(I_1-qI_2)}{(I_1+I_2)}\,.
\label{eq:rat1}
\ee
Moreover, considering the ratio of the coefficients from the CM boost and the CM drag, we have another ``predictor''
\be
\eta_2=\frac{q}{q+1}\frac{(I_1-qI_2)}{(I_1+q^2I_2)}.
\label{eq:rat2}
\ee
To summarize, in principle, binary configurations could experience a considerable kick if the CM boost and both the predictors $\eta_{1,2}$ are large enough. To combine the two features, in the lower panel of Fig.~\ref{fig:force} we show the product $\eta_1\,\eta_2$ as function of the velocity and mass-ratio. The combination of the two panels of Fig.~\ref{fig:force} can serve as a better reference to analyze potential kicks in binary configurations subject to dissipative forces.

In the rest of this work, we adopt units where $G=c=1$. In these units, the number $\rho M^2$ and $r/M$ are dimensionless. For future reference, these numbers are
\begin{align}
\frac{G^3}{c^6}\rho M^2&=1.6\times 10^{-18}\frac{\rho}{\rho_{\rm water}}\frac{M^2}{M_{\odot}^2}\,,\label{eq:density_normalization}\\
\frac{c^2 r}{G M}&=4.75 \times 10^5\frac{M_\odot}{M}\frac{r}{R_\odot}. \label{eq:conversion_m}
\end{align}
%

\section{Binary evolution in an environment}
\label{sec:num}
\subsection{Small-density expansion}\label{sec:smalld}

%
The set of equations~\eqref{eq:center}--\eqref{eq:disp} contains complex dynamics and is in general difficult to solve analytically.
To make some progress at the analytical level, consider a perturbative scenario in which the density of the environment is extremely small. In $G=c=1$, the solution can be perturbatively written as
\begin{align}
\mathbf{r}&=\mathbf{r}_0+\rho\mathbf{r}_\rho+{\cal O}(\rho^2),\\
\mathbf{R}&=\mathbf{R}_0+\rho\mathbf{R}_\rho+{\cal O}(\rho^2).
\end{align}
By expanding the equations, we find that (we chose that $\mathbf{R}_0=\mathbf{0}$, so the CM is initially at the center of the coordinate system)
\begin{equation}
\ddot{\mathbf{R}}_\rho=\frac{q M}{(q+1)^3} [I({v}_{0,1})-q I({ v}_{0,2})]\dot{\mathbf{r}}_0\,,
\end{equation}
where ${v}_{0,i}$ is the zeroth-order speed of the $i$-particle, being given by
\begin{equation}
{v}_{0,1}= \frac{\Omega  a q  }{(q +1)},~~~
{v}_{0,2}= \frac{\Omega a }{(q +1)},
\end{equation}
with $a$ being the orbital radius and $\Omega=\sqrt{M/a^3}$ the standard Keplerian frequency. The above equation can be solved analytically and the CM position at first order in the density in Cartesian coordinates is
\begin{align}
\mathbf{R}=\frac{\rho \,q\, a M}{(q +1)^3\Omega} [I({v}_{0,1})-q  I({v}_{0,2})]\times
\{\sin(\Omega t)-\Omega t,1-\cos(\Omega t)\}.\label{eq:low_density}
\end{align}

Note that the entire CM position is (anti-) symmetric to the substitution $q\to 1/q$, as it should. As expected, the CM position is zero for symmetric binaries or for a vanishing environmental density (when $\rho=0$). Figure~\ref{fig:analytical} compares the analytical prediction \eqref{eq:low_density} against a numerical evolution of the CM position and velocity for a mass ratio $q=2$ binary. The results are in excellent agreement in the initial stages of the evolution. As the orbit acquires eccentricity, the CM speed starts to oscillate with the same frequency as the epicyclic one, while increasing in its speed. The analytical approximation at this order only predicts the oscillation, with the speed-increase in this stage being encoded in higher-order terms and not captured by Eq.~\eqref{eq:low_density}.

Note that at late times, the CM velocity already reaches very high values, of order a few hundred kilometers per second. Eventually, the CM velocity goes down for very tight binaries, since it now behaves as a single body being dragged as it moves through a medium.
\subsection{CM kicks in uniform-density environments}
%
\begin{figure}
	\includegraphics[width=\linewidth]{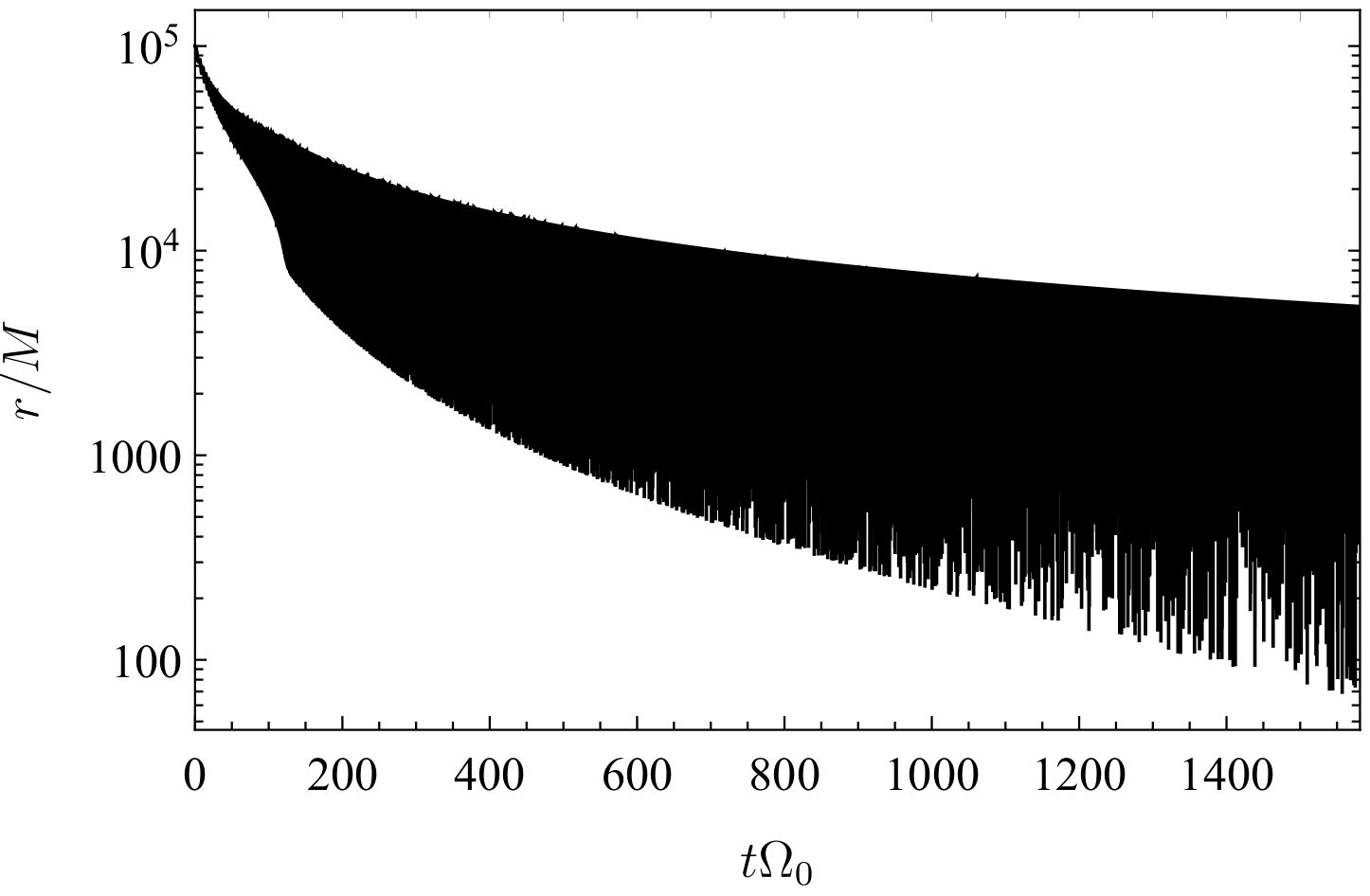}
	\caption{Radial distance as function of time, for a binary with initial separation of $a=10^5 M$, $\sigma=50{\rm km/s}$, and $q=2$. The density of the medium is $\rho M^2=10^{-20}$. The orbital distance decreases with time, with the orbit becoming more and more eccentric. In the final stages the eccentricity reaches values of the order $\sim 0.9999$. See the supplementary material for animations describing this specific orbit (also provided in \citet{animations}).}
	\label{fig:radius}
\end{figure} 
\begin{figure*}
	\includegraphics[width=1\linewidth]{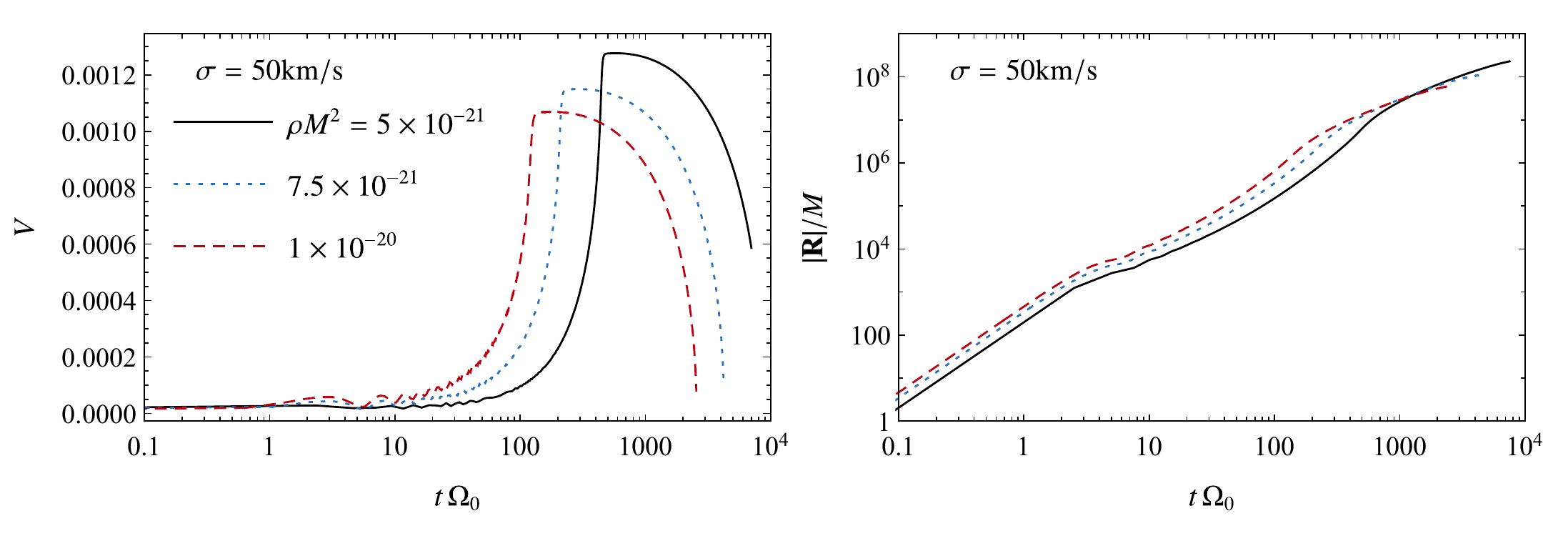}\\
	\includegraphics[width=1\linewidth]{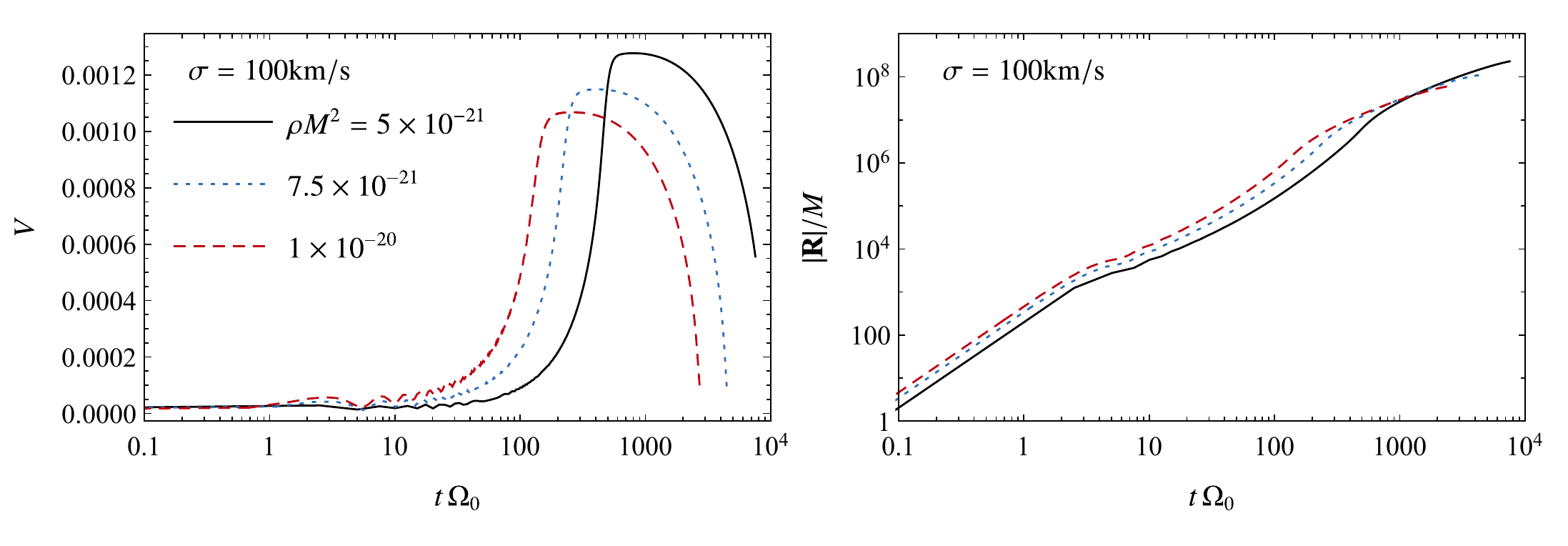}
	\caption{Velocity and displacement of the CM as function of time, for $\sigma=50\,{\rm km/s}$ (top panels) and $\sigma=100 {\rm km/s}$ (bottom panels), initial orbital separation of $a=10^{5}M$, and different environmental densities. All binaries have a mass ratio $q=2$.}
	\label{fig:kick_sig}
\end{figure*} 
In general, the differential equations given by \eqref{eq:center} and \eqref{eq:disp} need to be solved numerically. In the absence of dissipative forces, a known closed-form solution corresponds to circular orbits. As such, we use these ciruclar orbits to as initial conditions in our equations and monitor the system as it evolves acted upon by dynamical friction and the gravitational interaction. More details about the equations used in the numerical integrations and their initial conditions can be found in Appendix ~\ref{app:equations}.

We consider exclusively setups where the density is small enough that it takes some cycles before the CM acquires considerable speeds. Thus, the previous (closed-form, analytical) predictions for low-density environments should hold at early stages of the motion. To perform the integrations, we normalize all quantities in terms of the total mass of the system $M$. In this way, we are left with the space of parameters $(\rho,\,q,\,\sigma,\,a)$. 
The quantities $(\sigma,a)$ are related in the ``predictors", being described in Fig.~\ref{fig:force} by the vertical axis ($v/\sigma$), as the initial binary speed and the initial separation are related through $v=\sqrt{M/a}$.
Nonetheless, we will explore a variety of configurations, trying to verify whether the CM motion is viable in astrophysical situations.

Figure~\ref{fig:radius} illustrates the effects of dynamical friction on the binary, showing that the binary separation $\mathbf{r}(t)$ of what was an initially circular orbit starts varying wildly after a few cycles. After some cycles, the binary acquires a considerable eccentricity. The effect of increasing the eccentricity of the orbit under the influence of dynamical friction was also observed before by~\cite{Macedo:2013qea}. In the final stages of the evolution (when the CM speed decreases due to friction), the eccentric reaches values as high as $0.9999$. Some movies concerning this particular orbit binary are available in the supplemental material and also in \cite{animations}, and help visualizing the orbital evolution. We note here that even binaries with larger orbital separations and lower-density profiles behave in similar ways.

Figure~\ref{fig:kick_sig} illustrates the dependence of the CM velocity on the environmental density, showing the CM position and velocity for different densities.
Decreasing the medium density has the effect of \textit{increasing} the peak CM velocity. This is a cumulative effect: although the density and the CM force are smaller, the binary can go through many more cycles providing a large CM speed before CM drag slows it down. As a consequence, for smaller densities the peak CM velocity occurs at later stages in the binary evolution. We note that
we find CM speeds of the order of $10^{-3}c$, i.e., of order of $300 {\rm km/s}$, but they could be higher depending on the binary initial configurations and media density. The CM displacement can be orders of magnitude larger that the initial periastron. Note also that the peak CM velocity is attained at relatively large binary separation, c.f. Figs.~\ref{fig:radius} and \ref{fig:kick_sig}.

For example, suppose that the binary is evolving within a disk surrounding a supermassive black hole of mass $M_{\rm SMBH}$. 
For geometrically thin disks, such as those suitable for describing systems with accretion efficiency $10^{-2} \lesssim f_{\rm Edd}\lesssim 0.2$,
one can solve the equations describing the disk's structure exactly in Newtonian theory and in a steady-state regime~\citep{2002apa..book.....F,shakura_sunyaev}. Its density $\rho$ and height $H$ are: 
\begin{align}
\frac{\rho}{\rho_{\rm water}}\approx& 2\times 10^{-4} f_{\rm Edd}^{11/20}\left(\frac{0.1}{\alpha_v}\right)^{7/10} \nonumber\\
&\times\left(\frac{10^6 M_\odot}{M_{\rm SMBH}}\right)^{7/10} \left(\frac{10^3GM_{\rm SMBH}}{c^2r}\right)^{15/8}\,, \label{eq:rhoF}\\
\frac{c^2H}{GM_{\rm SMBH}}\approx &7 f_{\rm Edd}^{3/20}  \left(\frac{0.1}{\alpha_v}\right)^{1/10} \left(\frac{10^6 M_\odot}{M_{\rm SMBH}}\right)^{1/10}\nonumber\\ 
&\times\left(\frac{c^2r}{10^3GM_{\rm SMBH}}\right)^{9/8} \,,
\end{align}
where $\alpha_v$ is a viscosity parameter.
The CM displacement is always along the orbital angular momentum plane (for these non-spinning binaries, at lowest post-Newtonian order).
Thus, any binary whose orbital plane is mis-aligned with that of the disk can be potentially be
kicked out of the disk through this environmental effect, in relatively short timescales.

\begin{figure}
	\includegraphics[width=\linewidth]{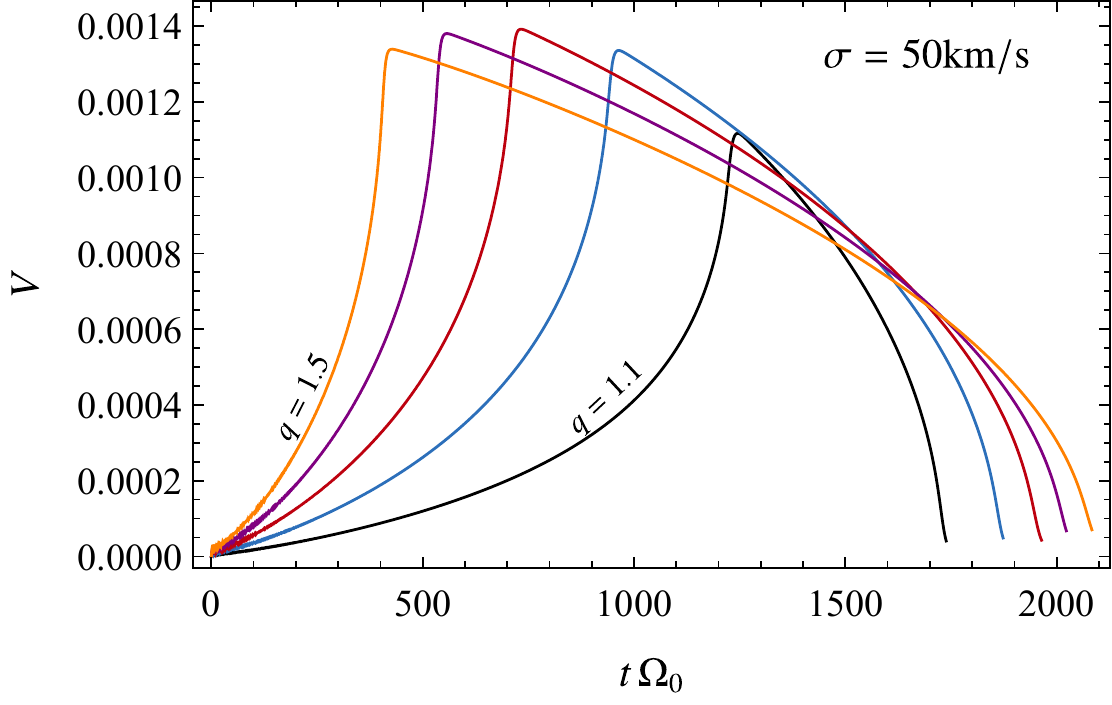}\\
	\includegraphics[width=\linewidth]{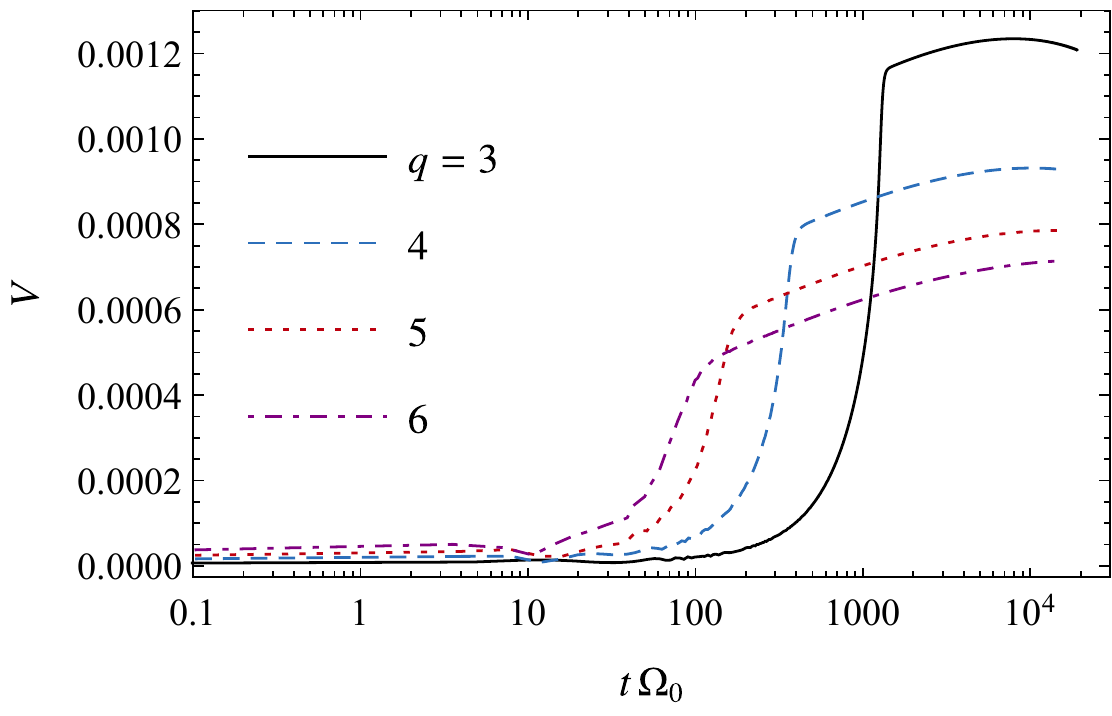}
	\caption{Dependence of the CM velocity on the mass ratio $q$. Top panel: Here we consider $\rho M^2=10^{-20}$, initial separation $a=10^{5}M$, and $\sigma=50{\rm km/s}$. (green markers in Fig.~\ref{fig:force}). For these particular initial conditions, the maximum CM speed happens for mass ratio $q\approx 1.3$.
		Bottom Panel: same for $\rho M^2=10^{-21}$, initial separation $a=10^{5}M$, and $\sigma=50 {\rm km/s}$. The curves refer to $q=3,4,5$ and $6$. (blue markers in Fig.~\ref{fig:force}).}
	\label{fig:vcmq}
\end{figure}

Having established the role of the density in the evolution of the binaries, we now investigate the influence of the other parameters, exploring the different regions depicted in Fig.~\ref{fig:force}. 
Let us fix $(a,\sigma)=(10^5M,50{\rm km/s})$, changing the mass-ratio (green dots in the left part of Fig.~\ref{fig:force}). Since fixing $a$ corresponds to fixing the initial speed, the dots belongs to the same horizontal line in the predictors plot. Fig.~\ref{fig:vcmq} shows the CM speed for different mass ratios. For symmetry reasons, the net CM velocity is zero both for equal-mass binaries ($q=1$) and for extreme-mass-ratio binaries ($q\to \infty$). 
As we argued, for nearly equal-mass binaries the CM ``kick'' is suppressed. Nonetheless, even for mass ratios $q=1.1$ the CM speed is still relevant and of order $10^{-3}c$. We can also see that the maximum speed seems to be maximized around $q\sim 1.3$. Note that the mass ratios explored in the figure are within the window of the ones observed by LIGO-VIRGO.

The lower panel of Fig.~\ref{fig:vcmq} shows the CM velocity dependence at higher mass ratios, cf. the rightmost (blue) dots in Fig.~\ref{fig:force}. 
Note that we consider smaller environmental densities now, such that the binary goes through more orbits before an appreciable boost builds up. As we increase the mass-ratio, the drag force becomes more important than the CM boost force and, therefore, the predictor tells us that the CM speed gradient should decrease earlier in the binary evolution. This is confirmed in the bottom panel of Fig.~\ref{fig:vcmq}.

\begin{figure}
	\includegraphics[width=\linewidth]{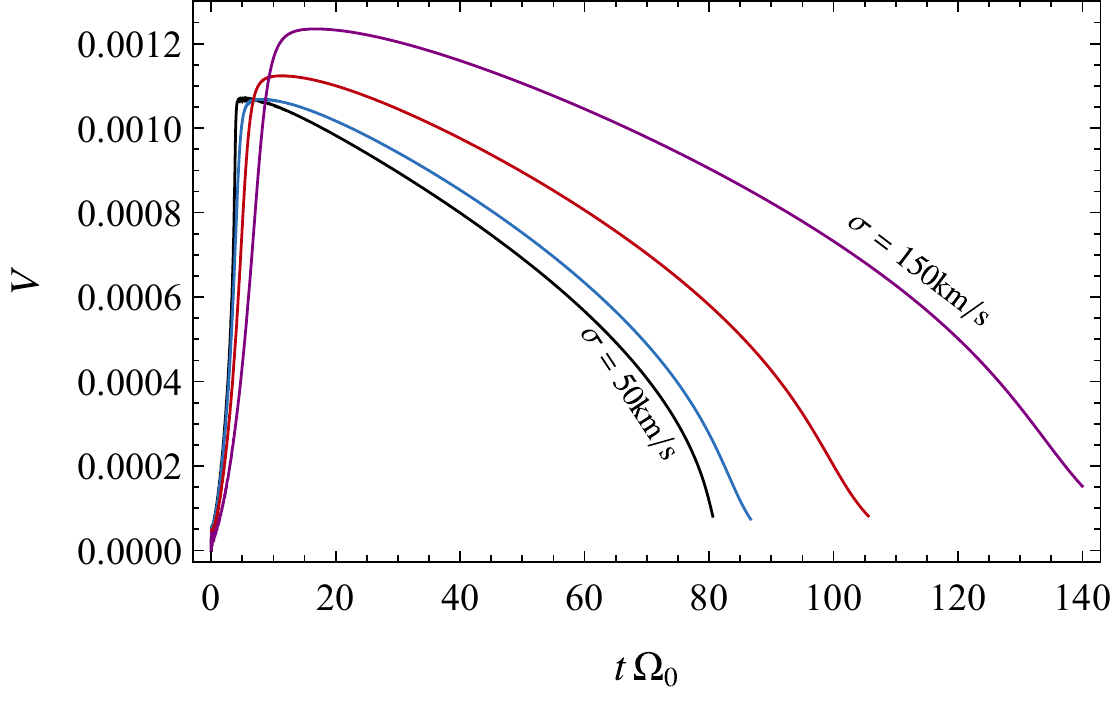}
	\caption{Dependence of the CM kick on $\sigma$. Here we consider $\rho M^2=10^{-20}$, initial separation $a=10^{5}M$, and $q=0$. The curves refer to $\sigma=50,75,100,125$ and $150{\rm km/s}$. (red markers in Fig.~\ref{fig:force})}
	\label{fig:kicks}
\end{figure}
\begin{figure}
	\includegraphics[width=\linewidth]{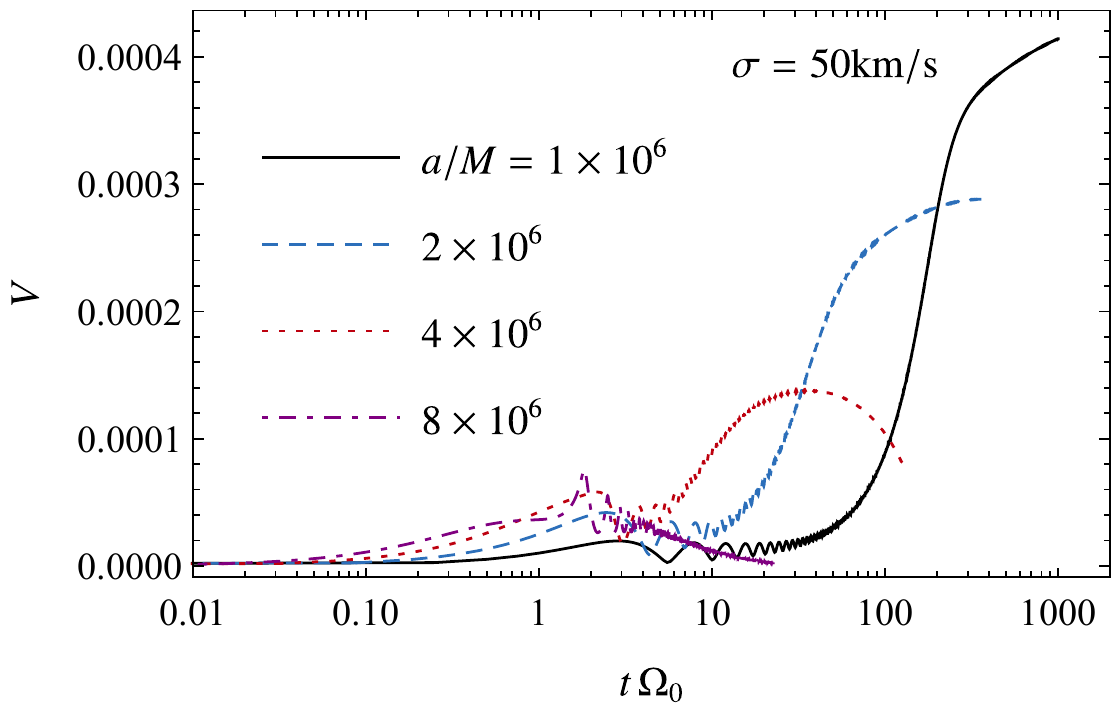}
	\caption{Dependence of the CM kick on the initial separation $a$. Here we consider $\rho M^2=10^{-23}$, mass-ratio $q=3$, and $\sigma=50{\rm km/s}$. (magenta markers in Fig.~\ref{fig:force}).}
	\label{fig:kicka2}
\end{figure}
Now we investigate the red dot configurations in Fig.~\ref{fig:force} (vertical dots on the left part of the plot). In Fig.~\ref{fig:kicks} we show the CM kick for medium density $\rho M^2=10^{-20}$, $q=2$, and different values of $\sigma$. Increasing the value of $\sigma$ makes the point move downwards in the predictors. We can see from Fig.~\ref{fig:kicks} that the maximum speed of the CM increases with $\sigma$ in this regime.

We can further investigate the dependence on the orbital separation by looking into a different regime. This is done considering the configurations on the vertical dots in the bottom part of Fig.~\ref{fig:force} (magenta dots). These are obtained by considering $\sigma=50{\rm km/s}$, $\rho=10^{-23}$ and different values of $a$. The predictors tell us that increasing the value of $a$ in that region, moving downwards, we should expect the kick to decrease. The result shown in Fig.~\ref{fig:kicka2} agrees with this prediction.

The above results show that the predictors in Fig.~\ref{fig:force} provide a powerful tool to investigate binaries with potential kicks. A more careful analysis, however, still has to be done in a case-by-case basis. With the examples provided here, we have shown that in many scenarios the CM boost can be quite significant. We also highlight the fact that the peak boost depends on the density of the medium. In the following, we further investigate the density dependence in some particular cases.

\subsection{Maximum CM speed and the density of the medium}
%
\begin{figure*}
	\includegraphics[width=\columnwidth]{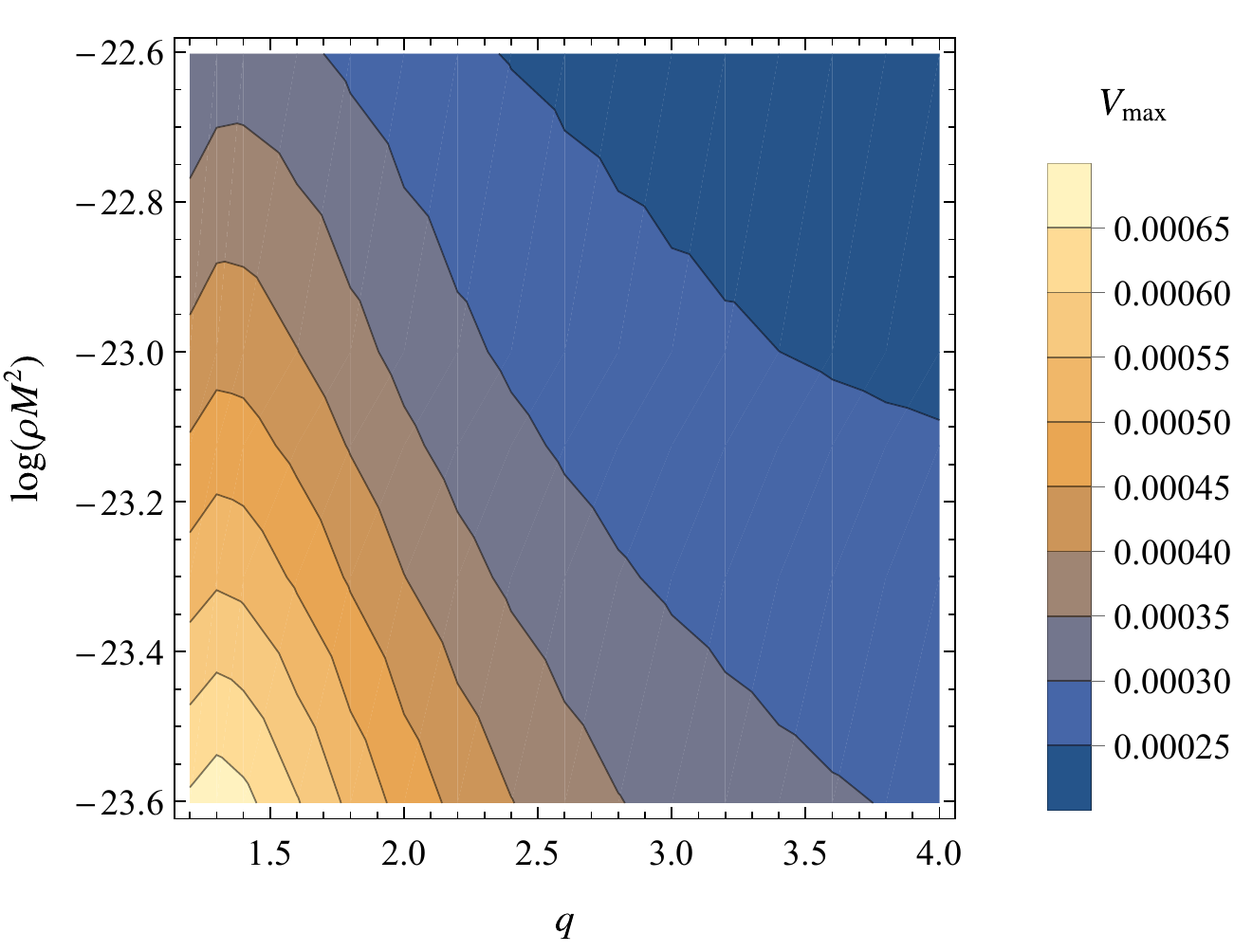}	\includegraphics[width=\columnwidth]{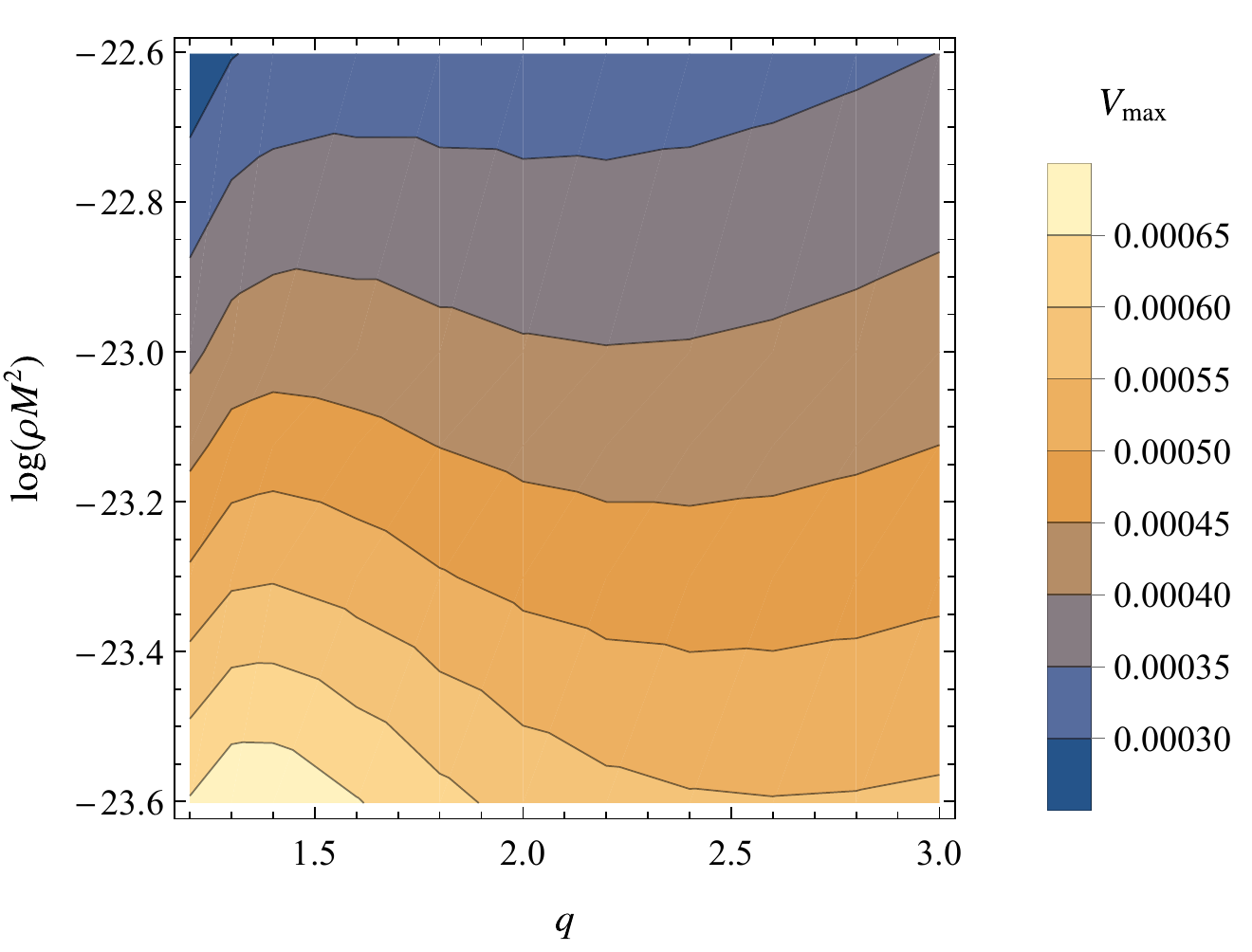}
	\includegraphics[width=\columnwidth]{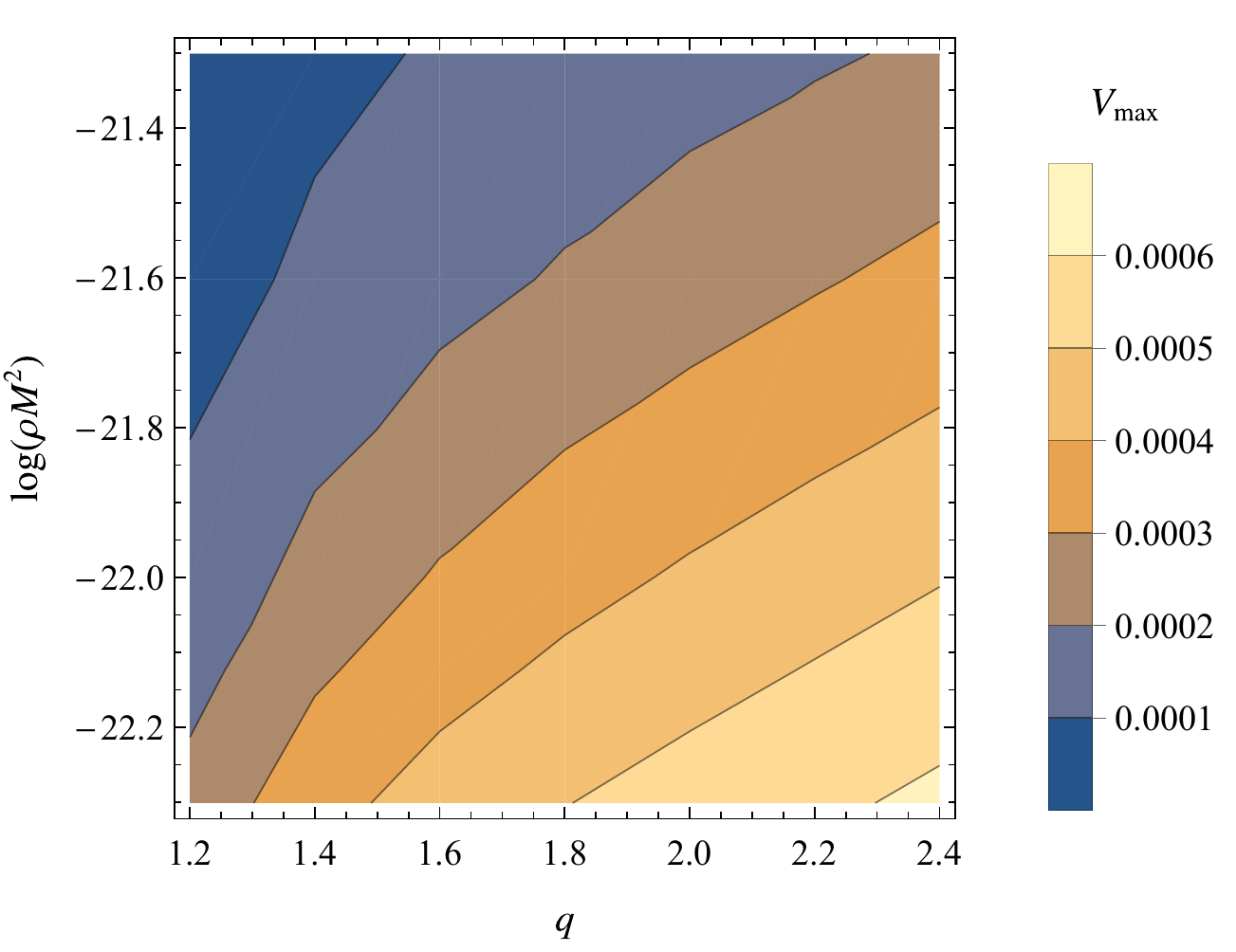}	\includegraphics[width=\columnwidth]{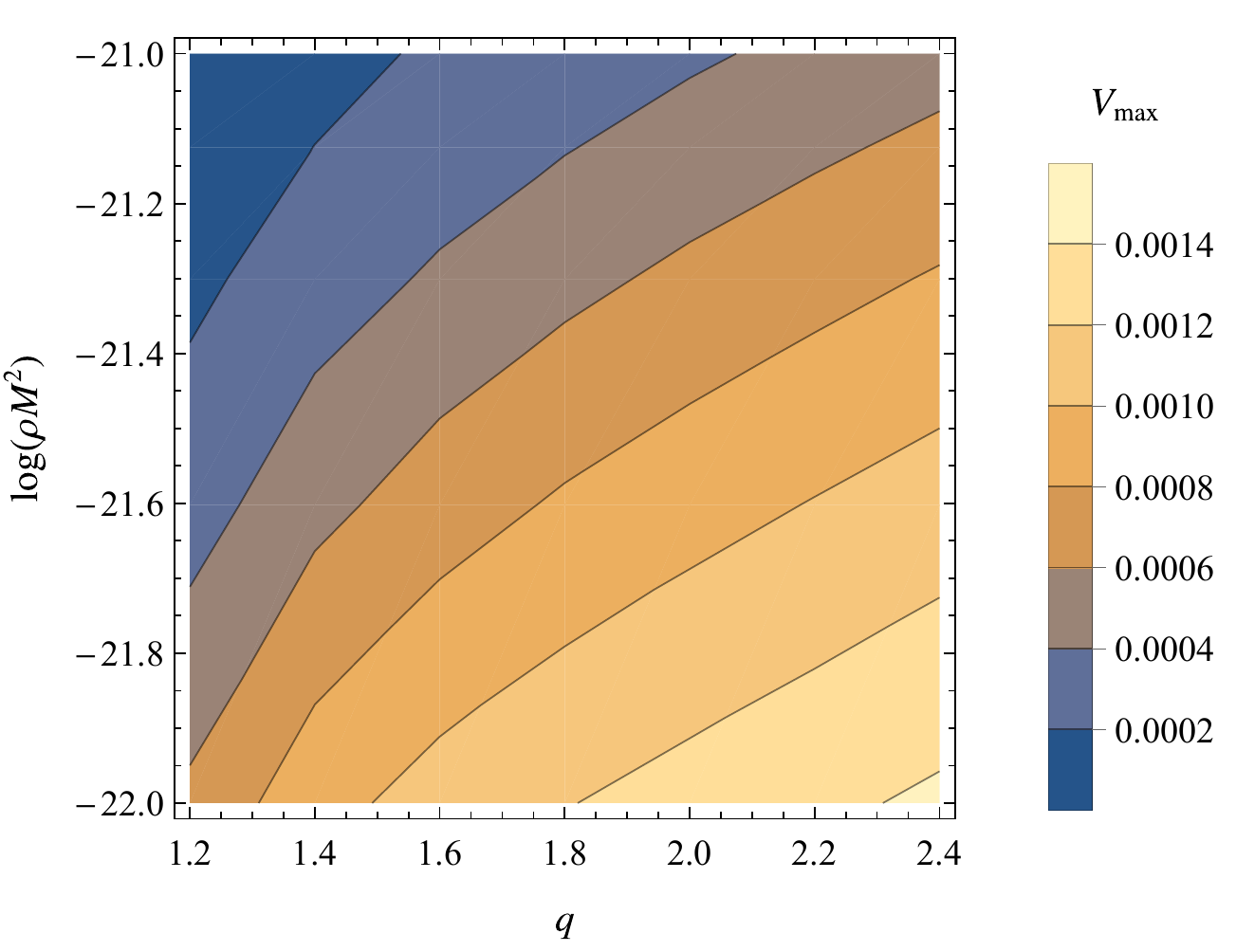}
	\caption{Maximum CM speed for $\sigma=25{\rm km/s}$ (top-left), $50{\rm km/s}$ (top-right), $100{\rm km/s}$ (bottom-left), and $200{\rm km/s}$ (bottom-right). Note that the maximum speed increases for lower-density media, as pointed before. Here we consider initial separation to be $a=10^6M$.} 
	\label{fig:contour_k}
\end{figure*}
As we discussed in the context of Fig.~\ref{fig:kick_sig}, the density of the medium plays a crucial role in the maximum speed acquired by the CM. It is, therefore, important to understand how $V_{\rm max}$ changes with the density within the parameter space for binary systems. In this subsection we focus on the particular case of a binary separation $a=10^6M$.
Fixing the initial orbital separation, we are still left with $(\sigma,\rho,q)$. As such, in order to map the maximum CM speed, we fix $\sigma$ to be $25{\rm km/s}$, $50{\rm km/s}$, $100{\rm km/s}$, and $200{\rm km/s}$, spanning over different ranges for the density $\rho$ and mass-ratio $q$. The configurations are represented in Fig.~\ref{fig:force} as horizontal solid lines.

Figure~\ref{fig:contour_k} summarizes our results. For a fixed value of the mass-ratio, $V_{\rm max}$ is larger for lower-density media, as noted previously when discussing Fig.~\ref{fig:kick_sig}. Note, however, that many more cycles are necessary for the CM boost in the lower-density region. In fact, it gets increasingly computationally hard to evolve such orbits.

The behavior of $V_{\rm max}$ for fixed and varying mass-ratio is less obvious. The behavior of $V_{\rm max}$ is sensitive to the value of $\sigma$, as can be noted comparing the different plots in Fig.~\ref{fig:contour_k}. Nonetheless, we expect that the CM speed eventually goes to zero for higher values of $q$, as the term proportional to $\dot{\mathbf{r}}$ in Eq. \eqref{eq:center} vanishes in that limit. In fact, the large-$q$ limit can be solved analytically, in a similar fashion as the one showed in Sec.~\ref{sec:smalld} for small-density expansion. Unfortunately, we were unable to evolve the binaries further in higher mass-ratio regime in the fully numerical setup.

\begin{figure}
	\includegraphics[width=\linewidth]{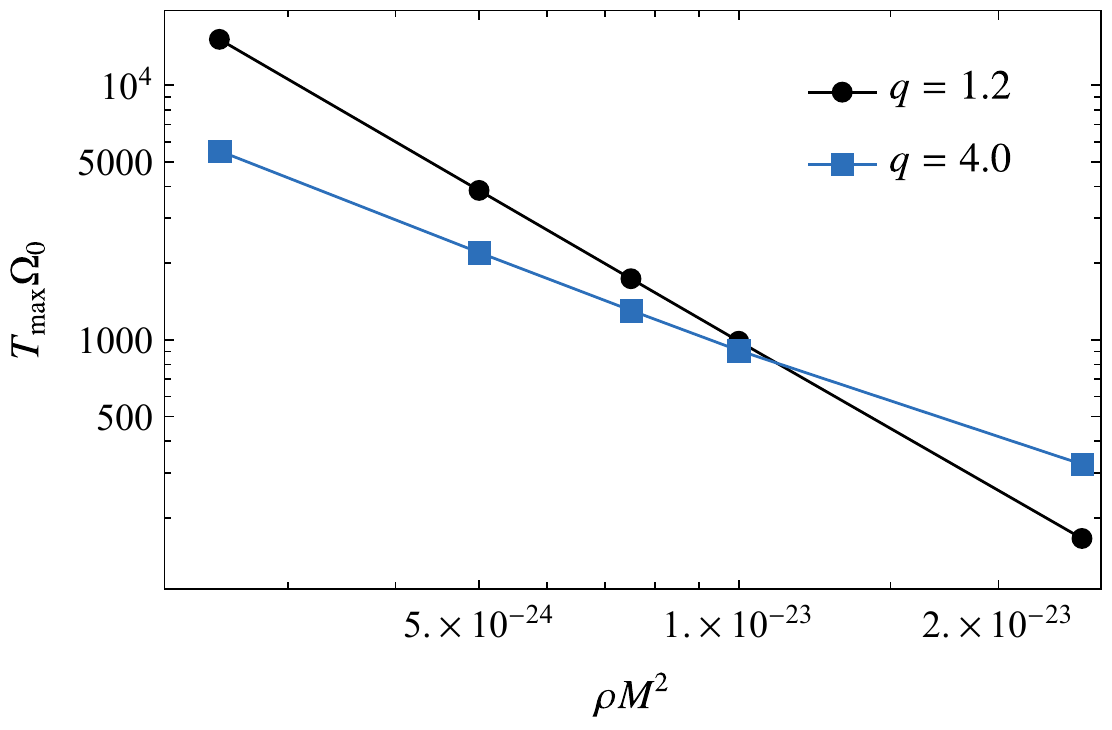}
	\caption{Time interval for the CM of binaries with initial separation $a=10^6M$ and $\sigma=25 {\rm km/s}$ to achieve maximum speed. A similar qualitative result holds for other values of $\sigma$.}
	\label{fig:time}
\end{figure}
Considering that lower-density media generate both higher-speeds and a longer evolution for binaries, one can search for which binaries the effect is relevant and on which timescales. This can be answered by analyzing the time $T_{\rm max}$ it takes for the binary to evolve up to the maximum CM speed. For simplicity, let us focus on the case $\sigma=25{\rm km/s}$ (top-left panel of Fig.~\ref{fig:contour_k}). The result is shown in Fig.~\ref{fig:time} for $q=1.2$ and $q=4.0$. Note that the behavior of $T_{\max}$ as function of $\rho$ seems to be a power-law in the interval analyzed, with the power being dependent on $q$. In particular, for $q=1.2$, a good description of our results is the expression
\begin{equation}
T_{\rm max}=6\left(\frac{M_\odot}{M}\right)^3\left(\frac{10^{-6}\rho_{\rm water}}{\rho}\right)^2 {\rm years}\,,\label{eq:times}
\end{equation}
which reinforces that the CM boost in astrophysical environments can be observable in a reasonable timescale. Notice that Eq.~\eqref{eq:times} was obtained in a small-density regime, such that the particle evolves through many orbits before the CM reaches the maximum speed. As such, we do not expect it to be valid for larger density profiles. Finally, we verified that expressions similar to \eqref{eq:times} still hold for other values of $(\sigma,q)$ explored in Fig.~\ref{fig:contour_k}, but with different numbers for the exponents of mass and density. This ensure us the feasibility of environmental kicks in a observable timescale in possible astrophysical scenarios.

\section{Conclusions}
Accretion disks, common-envelope systems or other nontrivial environments seed formation of compact binaries.
It is natural to wonder what effect does the environment have on the evolution of binaries.
We have shown that binaries are good ``swimmers,'' and move easily through their surroundings -- with the help of gravitational drag and accretion -- achieving large speeds (up to $300\, {\rm km/s}$ or more). Such an effect is easily measurable with Doppler shifts either in the EM or gravitational-wave band. It could also lead to the ejection of the binary off the medium where it evolves.

Despite the similarities with the common-envelope case, the numbers shown here should be taken with caution: inspiralling bodies will backreact on the environment, but the gravitational-drag 
results we used are oblivious to such an effect. We have also not included accretion or secular effects
in the evolution of the binary; although we argued that they are subdominant, a inclusion of additional physics is clearly necessary
(such as the ones presented in \cite{Passy_2011}, for example). As already pointed in the main text,
it would also be interesting to understand the influence of the interference between the wakes of the orbiting bodies, as noted by \cite{Kim:2008ab} and \cite{Baruteau:2010sm}. These wake interferences will certainly have some impact, especially when the binaries are at periastron. 
We have neglected third massive bodies in the problem; in particular, the effect
of the massive black hole harboring the accretion disk is not taken into account here.
The approximation of a continuous distribution of background point objects will break down when modeling a background of stars, planets or large bodies, and interactions with individual objects will become important. This was modeled by \cite{Leigh:2017wff}, who found that less than 10 such encounters was sufficient to harden binaries in nuclear star clusters (to the point that fast merger by gravitational-wave radiation could occur). This effect will further limit our $\eta_1$ discriminator, but is irrelevant for gas or plasma distributions.
From the theory side, a realistic model of binary gravitational drag which can be solved in closed-form would
clearly be interesting. All the additional physics above will have an impact in the predictors for CM boosts, but it is uncertain whether they could prevent such high speeds as the ones discussed here from being achieved. Note in particular that large CM velocities are possible even when the separation between the bodies is large. Our results are of course a part of previous studies with N-bodies or gravitational drag, but the possibility of transmitting large kicks seems to have gone unnoticed.

Very recently, a candidate electromagnetic counterpart to a binary black hole merger was discussed by~\cite{Graham:2020gwr}. This study argues that the binary, of total mass $\sim 100 M_{\odot}$, evolved and merged within an environment of gas density $\sim 10^{-10} {\rm g \,cm}^{-3}$. Comparison with Eq.~\eqref{eq:density_normalization}, tell us that 
for this system $\frac{G^3}{c^6}\rho M^2\sim 10^{-24}$, close to the values we studied. In other words, there is the tantalizing possibility that
the effect we just discussed may have interesting and observable consequences for gravitational-wave astronomy.

\section*{Acknowledgements}
%
V. C. would like to thank Waseda University for warm hospitality and support 
while this work was finalized. 
V. C. acknowledges financial support provided under the European Union's H2020 ERC 
Consolidator Grant ``Matter and strong-field gravity: New frontiers in Einstein's 
theory'' grant agreement no. MaGRaTh--646597. 
C.F.B.M. would like to thank Conselho Nacional de Desenvolvimento
Científico e Tecnológico (CNPq), and Coordenação de Aperfeiçoamento de Pessoal de Nível Superior (CAPES), from Brazil.
This project has received funding from the European Union's Horizon 2020 research and innovation 
programme under the Marie Sklodowska-Curie grant agreement No 690904.
We thank FCT for financial support through Project~No.~UIDB/00099/2020 and through grant PTDC/MAT-APL/30043/2017.
The authors would like to acknowledge networking support by the GWverse COST Action 
CA16104, ``Black holes, gravitational waves and fundamental physics.''
%







\appendix

\section{Differential equations for the orbital evolution in Cartesian coordinates}
\label{app:equations}

In order to integrate the differential equations, we write Eqs.~\eqref{eq:center} and \eqref{eq:disp} in cartesian coordinates, by defining,
\begin{equation}
\mathbf{R}=\{X(t),Y(t)\},~~\mathbf{r}=\{x(t),y(t)\}.
\end{equation}
Additionally, we write the velocity of each individual particles in terms of the CM and orbital separation vectors as (these are used in the dynamical friction terms)
\begin{equation}
\mathbf{v}_1=\mathbf{V}-\frac{q}{1+q}\mathbf{v},~~\mathbf{v}_2=\mathbf{V}+\frac{1}{1+q}\mathbf{v},
\end{equation}
with $\dot{\mathbf{r}}=\mathbf{v}$, $\dot{\mathbf{R}}=\mathbf{V}$, $\dot{\mathbf{r}}_1=\mathbf{v}_1$, and $\dot{\mathbf{r}}_2=\mathbf{v}_2$. By using these relations into Eqs.~\eqref{eq:center} and~\eqref{eq:disp}, we obtain the following system of equations
\begin{align}
\ddot{X}(t)&=F_1(\dot{X},\dot{Y},\dot{x},\dot{y},x,y),\\
\ddot{Y}(t)&=F_2(\dot{X},\dot{Y},\dot{x},\dot{y},x,y),\\
\ddot{x}(t)&=F_3(\dot{X},\dot{Y},\dot{x},\dot{y},x,y),\\
\ddot{y}(t)&=F_4(\dot{X},\dot{Y},\dot{x},\dot{y},x,y).
\end{align}
The explicit form of the functions $F_i$ are rather lengthy, so we provide them in the companion {\sc Mathematica} notebook provided as a supplementary material. The above equations are solved considering that in the absence of dissipative forces the solution would correspond to circular orbits with the center of mass placed at the origin of coordinate system. As such, we impose the following initial conditions
\begin{align}
X(0)&=Y(0)=y(0)=0,\\
\dot{X}(0)&=\dot{Y}(0)=\dot{x}(0)=0,\\
x(0)&=a,\dot{y}(0)=\sqrt{\frac{G M}{a}}.
\end{align}

We evolve the above system of equations subjected to the initial conditions without considering dissipative terms (effectively setting $\rho=0$), checking that indeed the solution corresponds to circular orbits. Accuracy and precision were tested to verify the stability of the numerical solutions in order to check the correctness of the results.

We also explored a ``polar'' coordinate system, defining $\mathbf{r}=r(t)\{\cos(\varphi_r),\sin(\varphi_r)\}$ and $\mathbf{R}=R(t)\{\cos(\varphi_R),\sin(\varphi_R)\}$. Although the equations seems to be simpler in these variables, the integrations actually takes longer to evolve. This appears to be related with the functions $I_i$, that are linked with the particles speeds rather to the $(r,R)$ variables. We remark, however, that it would be interesting to search for simpler ways to write the system of equations to further explore the parameter space.

\section{Self-propulsion considering collisional media}\label{app:collisional}
%
\begin{figure}
\includegraphics[width=\linewidth]{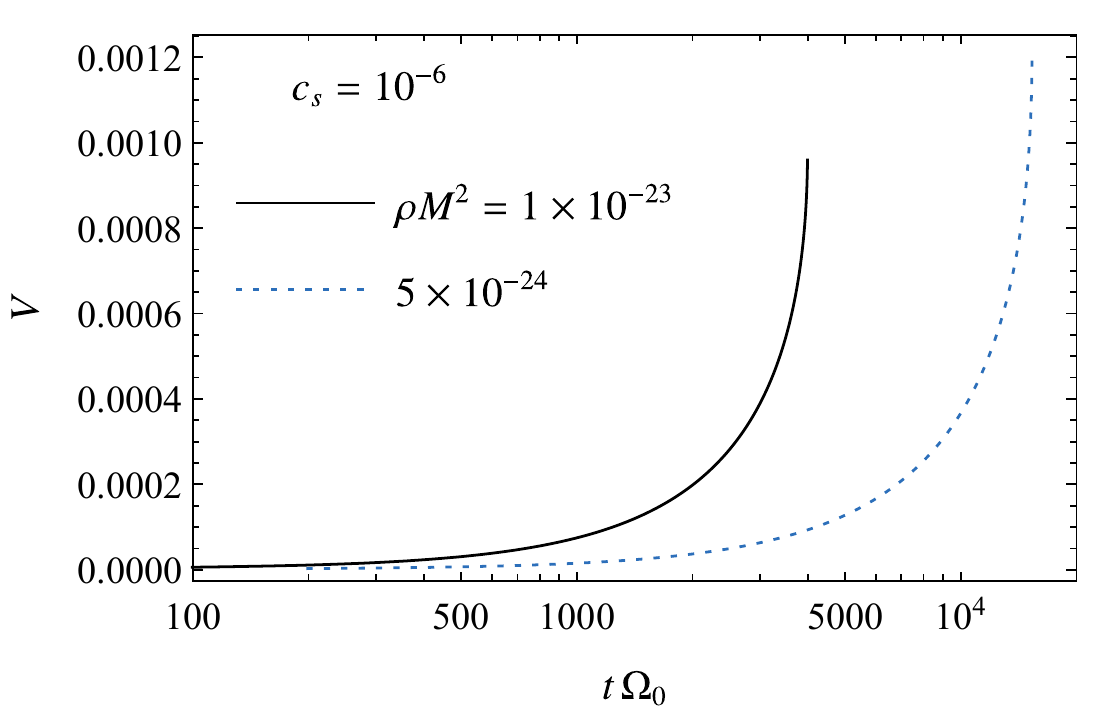}
\caption{CM speed as a function of time for a collisional fluid environment, considering the Ostriker model~\citep{Ostriker:1998fa}. We fix the speed of sound to be $10^{-6}$ (in units of the speed of light), mass-ratio $q=1.5$, and consider two values for the medium density.}\label{fig:vcm_col}
\end{figure}
In this appendix we provide some supplementary results, showing that high CM speeds can be reached also for binaries in a collisional fluid. We consider the Ostriker model, as presented in the main text [Eq.~\eqref{eq:dyn}], using $\Lambda=r_i(t)/m_i$ such that the formula describes well perturbers in circular orbits, as described by \cite{Kim:2007zb}. Since the transition between supersonic and subsonic is not smooth, we investigate only motion in the supersonic regime. However, we note that these transitions could further enhance the CM boost since they provide additional elements to the asymmetry between the $I_i$ functions.

In Fig.~\ref{fig:vcm_col} we show the CM speed as function of time for binaries with $q=1.5$ in media with density $\rho M^2=10^{-23}$ and $5\times10^{-24}$, with initial separation of $a=10^6M$. We consider $c_s=10^{-6}$, but we verified that the results are basically insensitive to $c_s$ at the interval $(10^{-4},10^{-6})$, changing only in the very late times before $m_2$ becoming subsonic. The figure shows that even in the scenario of collisional media the CM speed can reach hundreds of $\rm km/s$. We have stopped the integrations at the point in which $m_2$ becomes subsonic, as explained above.


\bsp	
\label{lastpage}

\end{document}